\documentclass[%
 reprint,
 amsmath,amssymb,
 aps,nofootinbib,   
]{revtex4-1}
\usepackage{bm}% bold math
\PassOptionsToPackage{linktocpage}{hyperref}
\usepackage[hyperindex,breaklinks]{hyperref}
\usepackage{enumitem}
\usepackage{slashed}

\renewcommand{\theta}{\vartheta}

\renewcommand{\vec}[1]{\ensuremath{\boldsymbol{#1}}}

\newcommand{\bra}[1]{\ensuremath{\left< #1\,\right|}}
\newcommand{\ket}[1]{\ensuremath{\left|\, #1\right>}}

\usepackage{array}
\usepackage{mathtools}

\usepackage{etoolbox}

\usepackage{xspace}

\newcommand*{\fig}{Fig.\@\xspace}

\usepackage{subcaption}

\begin{document} 

\title{Universe's Primordial Quantum Memories}

\author{Gia Dvali$^{1,2,3}$, Lukas Eisemann$^{1,2}$, Marco Michel$^{1,2}$, Sebastian Zell$^{1,2}$}
\affiliation{%
$^1$ 
Arnold Sommerfeld Center, Ludwig-Maximilians-Universit\"at, Theresienstra{\ss}e 37, 80333 M\"unchen, Germany, 
}%
 \affiliation{%
$^2$ 
Max-Planck-Institut f\"ur Physik, F\"ohringer Ring 6, 80805 M\"unchen, Germany
}%
 \affiliation{%
$^3$ 
Center for Cosmology and Particle Physics, Department of Physics, New York University, 726 Broadway, New York, NY 10003, USA
}%

\date{\today}

\begin{abstract}  
We provide a very general argument showing that the Universe must have kept its quantum memories 
from an epoch much earlier than $60$ e-foldings before the end of inflation. 
 The point is that a generic system of enhanced memory storage capacity exhibits a phenomenon of {\it memory burden}. 
  Due to its universal nature this effect must be applicable to 
de Sitter since the latter has a maximal memory storage capacity thanks to its Gibbons-Hawking entropy.    
The primordial information pattern encoded in de Sitter memory initially 
costs very little energy. However, because of Gibbons-Hawking evaporation, the memory burden of the pattern
grows in time and increasingly back reacts on the evaporation process.
After a finite time the memory burden becomes unbearable 
and de Sitter quantum breaks. 
 If inflation ended not long before its quantum break-time, the imprints of the primordial  memory pattern can be observable. This provides a qualitatively new type of window in the Universe's    
beginning, a sort of cosmic quantum hair. 
\end{abstract}

\maketitle
\section{Quantum Breaking of de Sitter} 
 De Sitter space  represents  a geometric description of the 
 cosmological state produced by a (nearly) constant 
 vacuum energy density $\Lambda$ of positive sign. 
   According to the idea of {\it inflationary cosmology} \cite{Guth}, 
   our present Universe underwent such a state at a very early stage
   of its cosmological evolution. Moreover, observations suggest 
  that we are again entering in a de Sitter like regime in the current epoch, albeit due to a minuscule value of vacuum-like energy density 
  $ \sim (10^{-3}eV)^4$ usually referred to as Dark Energy.   
  Thus, understanding the viability and properties of 
  de Sitter is of fundamental importance. 
 
  One thing we know for sure is that our Universe somehow found a graceful exit from the early de Sitter phase.  Within the inflationary paradigm this is achieved by a scalar field, the inflaton.  The latter acts as a classical clock guaranteeing that  the vacuum energy sooner or later is transformed 
  into a form of radiation after which the Universe evolves according 
  to the {\it Hot Big Bang}  cosmology. 
  
  It is customary to measure the duration of the de Sitter phase 
 in the number of e-foldings with each e-folding equal to the Hubble time
 $t_H = H^{-1} = {M_P \over \sqrt{\Lambda}}$, where $M_P$ is the Planck mass. How long did the de Sitter last?  
  In the standard inflationary picture the duration of the de Sitter phase is unbounded from above and could be arbitrarily long. 
 In such a picture, all the pre-existing information has been erased and the earliest data that can be obtained about our Universe is what has been created by quantum fluctuations approximately $60$ e-foldings prior to the graceful exit.  This is a rather sad prospect. \\
 
 However, the validity of the above perspective has been challenged in \cite{QB1,QB2,QB3}  
based on the following idea. 
 The well-known problem with de Sitter space is the impossibility of defining 
 an S-matrix vacuum. This creates a big challenge  
 since the S-matrix formulation is absolutely crucial for quantum gravity
 and string theory. 
 The idea of \cite{QB1} was to change the point of view: Instead of searching for a ``good" de Sitter vacuum
 we need to view the de Sitter itself as an excited state constructed on top 
 of a Minkowski vacuum. The latter must be taken as the true S-matrix vacuum.\footnote{Although an S-matrix can also be defined in AdS space, we shall 
 leave this space out of our discussion.}  
  The advantage of this new perspective  is that it 
 provides a well-defined framework  for  attempting to understand whether such a state exhibits some inconsistencies. 
  Thus, in this picture we must view de Sitter as a coherent state 
 $\ket{dS}$ of certain gravitational constituent quanta
  constructed on a Minkowski vacuum. 
  Requiring a maximally classical initial condition implies that the expectation values taken over this coherent state 
  must reproduce the classical metric picture. For example, the 
 Fourier harmonics of the classical metric of momenta $\vec{k}$ and 
 helicities $r$,  ($a_{\vec{k}}^{(r)},a_{\vec{k}}^{(r)*}$),
 are defined as the expectation values of the corresponding 
 creation/annihilation operators ($\hat{a}_{\vec{k}}^{(r)},\hat{a}_{\vec{k}}^{(r)\dagger}$), 
  \begin{equation} \label{VEV}  
  a_{\vec{k}}^{(r)} \rightarrow   \bra{dS} \hat{a}_{\vec{k}}^{(r)} \ket{dS} \,. 
  \end{equation}  
This fixes the occupation number and the characteristic frequencies
of the constituents of the state $\ket{dS}$ describing a de Sitter Hubble patch to be 
$N = {M_P^2 \over H^2}$ and $\epsilon_0 = H$ respectively. 
Note that the occupation number $N$ discovered in this way
incidentally comes out to be equal to the Gibbons-Hawking entropy  of the Hubble patch $S = {M_P^2 \over H^2}$.  
 
 The quantum resolution of de Sitter as a coherent state 
 constructed on a Minkowski vacuum sheds a very different light on some previously well-known phenomena.  Namely, as shown in \cite{QB1,QB2,QB3}, the rescattering  of the constituents 
 of the coherent state leads to its {\it decay}  via 
particle creation.  To leading order in the ${1\over N}$-expansion,  
this process reproduces the famous Gibbons-Hawking radiation \cite{GH}.
 However, since in this picture de Sitter is no longer a {\it vacuum} but a coherent state constructed on Minkowski, the Gibbons-Hawking particle creation
 is no longer a vacuum process.  Rather it is an emergent effective description of a more fundamental {\it Hamiltonian process} 
 of a {\it decay}  during which the constituents of the de Sitter coherent state
 get converted into free quanta. This reveals an 
 inevitable back reaction by which the particle creation changes
 the de Sitter state. Thus, during the Gibbons-Hawking particle creation 
de Sitter literally loses its quantum constituents and ``wears out". \\
 
 One of the key results obtained in this picture is that de Sitter exhibits 
 a phenomenon of {\it quantum breaking} \cite{qbreaking} beyond which 
 the true quantum evolution can no longer be matched by any 
 sensible classical dynamics. 
 An absolute gravitational upper bound on the quantum break-time of de Sitter is given by
 \cite{QB1,QB2,QB3}  
 \begin{equation}\label{Qt}
 t_Q = {1 \over n_{sp}} N t_H\,,
 \end{equation} 
 where 
 $n_{sp}$ is a number of active particle species \cite{QB3}. 
 Note that glimpses of similar time scales 
 can be seen already in semi-classical infrared effects on a de Sitter vacuum \cite{IR,measure}. However, the treatment of de Sitter as a coherent state on Minkowski vacuum reveals the underlying meaning of $t_Q$ as quantum 
 break-time. 

 Notice that the validity of the usual semi-classical assumption of zero back reaction 
 is recovered in the  limit  $N\rightarrow \infty$ with fixed background geometry. 
 Indeed, in this limit 
 the quantum break-time becomes infinite and de Sitter becomes  eternal. 
 The physical meaning of this limit  is easy to understand since it requires taking $M_P\rightarrow \infty$  while keeping $H$ finite. Thus, the vacuum energy density $\Lambda$ is taken to be infinite in this limit.  Not surprisingly, in such a case de Sitter can keep emitting particles at a constant rate for an infinite time. 
 Obviously, this cannot be achieved in any realistic 
 cosmological model.  In reality $N$ is finite and ${1 \over N}$-effects 
give important corrections over time scales of order $N$.  
 \\
 
  Two things are important to keep in mind. First, the quantum breakdown 
 is {\it real} and not an artefact of any sort of invalidity of perturbation theory. Secondly, it is {\it not } accompanied by any type of Lyapunov exponent or instability.     
  These two facts show that we are likely dealing with a genuine consistency bound indicating that in any consistent cosmology 
a de Sitter phase must end before its quantum breaking.  
  As explained in \cite{DGZ}, the latter constraint 
fully matches the bounds of the recently proposed \cite{swamp} 
``de Sitter Swampland conjecture". 
   How far we should take the above quantum breaking criterion  
 may be a matter of a dispute in which 
 we shall not enter in this paper. It is evident, however, that quantum breaking 
 of de Sitter is a physical phenomenon and requires to be fully understood.
 Due to the fundamental importance of the issue, any effective modeling 
 that can bring us one step closer to this understanding must be taken seriously.  \\
 
 \section{De Sitter Memory Burden}

  In \cite{QB1} it  has already been pointed out that ${1 \over N}$-effects
provide a {\it quantum clock} that encodes information about 
an actual duration of the de Sitter phase which can potentially give 
a new type of observable data from the inflationary epoch much before the last $60$ e-foldings.
Indeed, as soon as it becomes clear that de Sitter is subject to a quantum 
decay, its age becomes a physical observable. For example, density 
perturbations produced at different epochs will differ. This difference 
has nothing to do with the standard time variation of 
the Hubble parameter due to a classical slow-roll of an inflaton. 
Instead, it comes from the fact that the de Sitter background {\it ages}  due to 
its quantum decay. In other words, the back reaction that is measured by ${1 \over N}$-effects violates the de Sitter invariance in the same way 
as the evaporation of water from a finite volume tank violates the time-translation invariance. Even if the rate of the process is constant, the water 
level in the tank changes and this is an observable effect.  \\

 In the present paper we shall identify another effect that 
also encodes physical information about the duration of  the  de Sitter phase. 
This new observable can literally be viewed as the primordial quantum 
memory of de Sitter.   
 Namely, we shall argue that due to its Gibbons-Hawking 
 micro-state entropy the de Sitter state must be subject 
 to a universal phenomenon of {\it memory burden} \cite{Gia1} 
  that is generic for the systems with {\it enhanced 
 memory storage capacity}.  The latter term refers to systems that possess an exponentially large number of degenerate
  micro-states since such states can store quantum information patterns
  at a very low energy cost.  This capacity can be measured by the 
  associated micro-state entropy, the log of the number of degenerate 
  states.  
  By all counts, the de Sitter space must be a prominent member of the above category because of its 
  Gibbons-Hawking  entropy $S$. The latter, similarly to
 entropy of a black hole \cite{Bek}, saturates the Bekenstein 
 bound \cite{BBound} on information storage capacity.\\

 The essence of the phenomenon is the following. 
A state of enhanced memory capacity is achieved when  a certain macro-parameter $Y_0$ assumes a particular critical value $N$ for 
 which a set of modes with occupation number operators $\hat{Y}_k$, $k=1,2,...,S$, becomes gapless
 (where $S$ is some large number). 
 This behavior is parameterized by an effective energy gap function 
 ${\mathcal E}_k\left ({Y_0\over N}  \right )$ of the $k$-th mode in the Hamiltonian 
\begin{equation} \label{GapH} 
\hat{H} = \sum_k{\mathcal E}_k \hat{Y}_k\,,
\end{equation} 
 which vanishes 
 for $Y_0 \rightarrow N$ but is non-zero otherwise.  
 At the critical point $Y_0=N$, the modes $\hat{Y}_k$
 can therefore be excited at zero or very little energy cost.   Thus, an exponentially  
 large number of information patterns can be stored in the orthogonal micro-states that represent different 
number eigenstates of the $\hat{Y}_k$-modes, 
  \begin{equation} \label{Pattern}
 \ket{pattern}_{Y}  \equiv  \ket{Y_1,Y_2,...,Y_S}  \,.
  \end{equation} 
%$\ket{Y_1, Y_2,.....Y_S}$. 
Following \cite{Gia1}, we shall call the corresponding Fock space the {\it memory space}. Its dimensionality scales exponentially with $S$. 
Moreover, we will refer to the $\hat{Y}_k$-s as {\it memory modes}. 
Since these modes are gapless, the patterns of the sort (\ref{Pattern}) 
are nearly degenerate in energy. The corresponding memory storage capacity is consequently quantified by the micro-state entropy that 
 scales as $S$. \\
 
 In this situation,  the phenomenon of memory burden occurs \cite{Gia1}. Namely,
   any decay process that changes the value 
  of $Y_0$ takes the system away from a given critical state. 
 As a result the former modes are no longer gapless and 
 the pattern stored in them becomes very costly in energy. 
 This creates a {\it memory burden}  that back reacts on the decay process and tries to shut it down. \\   

  In this paper would like to suggest that because of the general nature of this phenomenon, it must also be applicable to de Sitter.    
   This mechanism reveals the existence of an alternative 
   {\it quantum clock} in the form of a memory burden of a  primordial pattern
   which we can denote as {\it M-Pattern}.  
  This pattern is encoded in the memory of nearly gapless modes that are responsible for the Gibbons-Hawking entropy. Due to the universality of the phenomenon, we do not need to know the precise microscopic origin 
 of these degrees of freedom. It suffices to know that they exist.  
 The classical evolution cannot affect the information stored in them.   
 Therefore the pattern
  cannot be erased by inflation and is revealed only after a long time
  due to cumulative quantum effects.  
  
   This phenomenon is intrinsically quantum in nature and 
   admits no classical or semi-classical understanding.  
   So past the point when the memory burden becomes unbearable, 
no classically viable description is possible. 
  Our interpretation is that the {\it memory overburden} 
  effect is an accompanying quantum informatics characteristics of 
  the phenomenon of {\it quantum breaking} of de Sitter described in \cite{QB1,QB2,QB3}.  \\
 
Besides its importance for general understanding, this  mechanism can open up a new observational window into a pre-inflationary Universe.   
Indeed, the closer the end of inflation is to its quantum breaking, the stronger 
 the influence of the quantum memory pattern becomes.  
 Thus, the situation with {\it M-pattern}  is exactly the opposite
 of other forms of pre-existing information which are readily eliminated by the subsequent de Sitter phase. 
 This creates a new prospect to search for the imprints of the primordial {\it M-pattern} in observational 
 data.  \\

 \section{Essence of Memory Burden Effect}  

We shall first briefly introduce the memory burden phenomenon studied in  
\cite{Gia1} and then apply it to de Sitter in the next section. 
We consider a generic quantum system consisting of two 
types of degrees of freedom which we shall describe 
by two sets of creation/annihilation operators 
$\hat{a}_k^{\dagger}, \hat{a}_k$ and $\hat{b}_k^{\dagger}, \hat{b}_k$, 
where  $k = 0,1,2,...,S$ are the labels.  The    
$a$ and $b$ sectors commute with each other and each satisfies the usual  algebra, 
  $  [\hat{a}_j,\hat{a}_k^{\dagger}] = \delta_{jk}\,, \, \, 
  [\hat{a}_j,\hat{a}_k]  =   [\hat{a}_j^{\dagger},\hat{a}_k^{\dagger}] =0\,$
 and the same for $b$-modes.     
The number operators are denoted by $\hat{Y}_k \equiv \hat{a}_k^{\dagger}\hat{a}_k$ and   $\hat{X}_k \equiv \hat{b}_k^{\dagger}\hat{b}_k$
respectively. 

 Let us first ignore the interaction among the sectors and 
give the simplest example of a Hamiltonian that has a poor memory capacity 
in the vacuum, but enormously enhanced memory capacity around some macroscopically excited state \cite{Gia2}, 
 \begin{eqnarray}  \label{H111} 
 &&\hat{H}_Y  = \sum_{k\neq 0} \epsilon_k \left ( 1 -  
 {\hat{Y}_0\over N}  \right )\hat{Y}_k\, 
  +  \,\epsilon_0  \hat{Y}_0 \,, 
  \end{eqnarray} 
 where  $N \gg 1$ is a large parameter. The  quantities $\epsilon_k$ represent the threshold excitation energies in the absence of 
 interactions.  The precise form of the spectrum is unimportant but 
 it is usual for quantum field theoretic systems that 
 the number of  
 modes increases with $\epsilon_k$. 
  We shall not assume any specific system and for us $k$ is just a label.  
  Throughout the paper we shall denote the expectation values by the same 
 symbols as the operators but without hats, e.g., 
 $\langle \hat{Y}_k \rangle = Y_k$. 
 In (\ref{H111})  we have singled out $\hat{Y}_0$ as a {\it master}  mode
 and ignored the interactions among the rest. 
 Their addition is trivial and changes nothing in the essence of the phenomenon (see, \cite{Gia2,Gia1}).  \\
 
In order for the reader not to get confused by our engineering 
efforts, we would like to clarify the following. 
Our goal by no means is to argue that states of enhanced memory capacity are generic.   They are in fact extremely 
special.  Our goal is to show  that once established by whatever 
means such a state dramatically influences the time evolution of the system. 
Therefore, it suffices to consider the simplest case. \\

  The memory patterns can be stored in the number eigenstate 
 ket vectors (\ref{Pattern}). 
   As in \cite{Gia2,Gia1}, we shall quantify the memory storage capacity of the system by the number of distinct patterns that fit within some fixed 
  microscopic gap.  For example, this can be chosen as $\epsilon_0$.   
 Consider first the patterns built around the $Y_0$-vacuum, i.e., when $Y_0 \ll N$. Unless $\epsilon_k$-s are very special, the
patterns cost  a lot of energy and are separated by a large energy gap. 
For example, the energy difference between the two patterns 
$Y_k=n_k$ and $Y_k = n_k'$ is given by $\Delta E = \sum_{k\neq 0} \epsilon_k
(n_k - n_k')$.    
  
However, as already shown in \cite{Gia2,Gia1}, the above 
system also delivers states of sharply enhanced memory storage capacity.    
Indeed, a nonzero expectation value of the master mode $Y_0$ 
lowers the effective energy thresholds of the modes $k\neq 0$ 
which are given by, 
\begin{equation} \label{eff}
{\mathcal E}_k = \epsilon_k \left ( 1 -  
 {Y_0\over N}  \right )\,.
\end{equation} 
They collapse to zero for a critical value $Y_0 = N$. 
The effect can be called an {\it assisted gaplessness}  \cite{Atoms} 
since the master mode {\it assists} other modes in becoming 
 gapless. 
The resulting macroscopic state has a sharply enhanced memory storage capacity 
as a large number of information patterns of the form (\ref{Pattern}) 
can be stored
at very low energy cost. 

Of course, in each particular case   the states 
will split into various super-selection sectors according to the symmetries of the system.  For example, if all the other interactions in Hamiltonian conserve the total occupation number
(such as, e.g., in the case of non-relativistic cold bosons 
of the type discussed in \cite{Gia1} or in \cite{qbreaking, Atoms})  
 the memory space must be chosen accordingly.

Although the states $\ket{Y_1,....,Y_S}$ are nearly-degenerate in energy, they exert different 
back reactions on the master mode $Y_0$ for different values $Y_{k \neq 0}$. This back reaction is measured 
by a quantity that we shall refer to as {\it memory burden} \cite{Gia1}. 
A general definition of it is: 
 \begin{equation}\label{MB}
 \mu \equiv \sum_{k\neq 0} Y_k{\partial {\mathcal E}_k \over \partial Y_0} \,.
 \end{equation}
 For the particular case of (\ref{eff}), this gives 
 \begin{equation}\label{burden}
 \mu \equiv - {\epsilon_{pat} \over N}\,,  
 \end{equation}
 where $\epsilon_{pat} = \sum_{k\neq 0} \epsilon_kY_k$ represents the
would-be cost of the pattern in the state $Y_0=0$, i.e., 
$\epsilon_{pat} $ is an unactualized energy 
cost.  It is important not to confuse this quantity with an actual energy cost of the pattern $E_{pat} = \langle \hat{H} \rangle$.

We will now investigate how the memory burden influences the decay 
process of the state of enhanced memory capacity into some external modes that themselves are not  
at the enhanced memory point.  The role of such modes will be played by 
$\hat{X}_k$.

   We shall introduce the simplest possible interaction that 
  allows the particle number transfer between the two  sectors. 
  It will be obvious that other choices do not affect the outcome.   
  Thus, we consider the following Hamiltonian:  
 \begin{eqnarray}  \label{H333} 
 &&\hat{H} =   \left(1 - {\hat{Y}_0\over  N} \right ) 
 \sum_{k \neq 0} \epsilon_k \hat{Y}_k \, 
   +  \\ \nonumber 
 && +  \epsilon_0  \hat{Y}_0 +  \sum_{k}  \epsilon_k \hat{X}_k \, 
   +  \\ \nonumber 
&& + \, {1 \over 2N} \sum_k \,\epsilon_k (\hat{a}_k^{\dagger} \hat{b}_k + \hat{b}_k^{\dagger}\hat{a}_k) \,  + \, ... \,.
 \end{eqnarray}
  The coefficient of the last term was normalized in such a way that 
  for $\mu =0$ the  state $Y_0=N$ would completely decay within the time 
 $\sim N\epsilon_0^{-1}$, i.e., it would lose on average one 
 $Y_0$-quantum per time $\sim \epsilon_0^{-1}$. 
 
  Next, we time evolve the system from the following initial state 
    \begin{eqnarray} \label{INstate}
 && \ket{in} = \ket{N,Y_1,...,Y_S}_Y \otimes \ket{0,0,...,0}_X \,. 
   \end{eqnarray} 

 The resulting occupation numbers as functions of time 
 are given by \cite{Gia1} 
  \begin{eqnarray} \label{evolve0} 
 && X_0(t) =  N A\, {\rm sin}^2\left ( {t \over \tau} \right) \, , \nonumber  \\
 && Y_0(t) =  N - X_0(t) \,,
 \end{eqnarray} 
 where 
  $A \equiv { 1 \over  1 + \left( {N\mu \over \epsilon_0 } \right )^2}$
 and  
 $\tau \equiv  {2 N\over \epsilon_0}  \sqrt{A}$. 
 These results illustrate the essence of the {\it memory burden} effect
 since the time evolutions for small and large values of $\mu$ are very different.  The critical value is set by $|\mu| \sim {\epsilon_0 \over N}$.
 
  Indeed, for  $|\mu| \ll {\epsilon_0 \over N}$ 
 we have $A \simeq 1$ and the occupation number of the master
 mode $Y_0$  almost fully diminishes after the time 
 $t \simeq {\pi N \over \epsilon_0}$.

 In contrast, in the case $|\mu| \gg {\epsilon_0 \over N}$
 the master mode only loses the following small fraction of its initial occupation number,  
\begin{equation} \label{frac}
 {\delta Y_0 \over Y_0} =  {\epsilon_0^2 \over 
 N^2 \mu^2} = \left (\epsilon_0 \over  \epsilon_{pat} \right)^2  \,. 
\end{equation}
   We see that the system is stabilized against the decay by the burden of its own memory. 
 
  The physics behind this phenomenon is very transparent.
  % and is due to  an interplay of the following two tendencies.  First, 
  When the system decays, it inevitably moves away from the critical state 
  of enhanced memory capacity. The stored memory pattern
  then becomes very expensive in energy because the memory modes 
  are no longer gapless.
  Indeed,  a decrease of the master mode by $\delta Y_0$ 
   increases the actual energy cost of the pattern by $\delta E_{pat} = \delta Y_0 |\mu|$.
    
   The presumed way out would be to offload the expensive pattern 
   into the $X$-sector together with the emitted $X_0$-particles. However, this is impossible due to an enormous energy splitting between the 
   $Y_k$-modes and their $X_k$-partners.  Recall that the later modes 
   are ``normal".  
   All this is very clear from the explicit form of time evolutions
  which for initial times takes the form,
    \begin{eqnarray} \label{evolveK} 
 && X_k(t) \simeq 
  {Y_k \over N^2} \, {\rm sin}^2\left ( {t \epsilon_k \over 2} \right) \,,
 \nonumber  \\
 && Y_k(t) = Y_k - X_k(t)  \, .
 \end{eqnarray} 
This shows 
  that  the pattern stored in $Y_k$-modes gets imprinted  
 into the corresponding $X_k$-modes with $1\over N^2$-suppressed 
 coefficients: 
 \begin{equation}  \label{copy} 
  \ket{pattern}_X =   \ket{{Y_1 \over N^2},...,{Y_S \over N^2}}_X \, .
 \end{equation}   
 Note that we are dealing with an intrinsically quantum effect due to finite 
 $N$.  \\

  After having explained the essence of the memory burden phenomenon, 
 we need to understand how different systems handle it.  The 
 following  two aspects are important. 
 The first one is the form of the functional dependence of $\mu$ on the 
control parameter $Y_0$ in the vicinity of a given enhanced memory state
$Y_0=N$.   
The second one is whether the memory enhancement takes place 
for some other values $Y_0 = N'$ and with what intervals.

  In the model (\ref{H111}) we assumed a simple linear dependence of 
 ${\mathcal E}_k$ on $Y_0$. In reality the dependence may be 
 non-linear. For example, we can take  
 \begin{equation} \label{Hgeneral}
 \hat{H} = \left (1 -  
 {\hat{Y}_0\over N}  \right )^m \sum_{k\neq 0} \epsilon_k \hat{Y}_k\,  + ...\,, 
 \end{equation} 
 with $m > 1$.
According to (\ref{MB}), the memory burden correspondingly depends on the departure from the critical state, $\delta Y_0 =
 N- Y_0$, as  
 \begin{equation} \label{mueff}
 \mu = - m \left ({\delta Y_0 \over N} \right )^{m-1} {\epsilon_{pat} \over N}\,. 
 \end{equation}
   In such a case the memory burden is a higher order effect 
 in ${\delta Y_0 \over N}$  
 and the  back reaction is delayed. 
 Equating the above to a critical value $\mu = -{\epsilon_0 \over N}$, we get an
 upper bound on $\delta Y_0$ above which the back reaction 
 from the memory burden cannot be ignored:
 \begin{equation} \label{dcrit}
  \delta Y_0  =  \left (\epsilon_0 \over  m\epsilon_{pat} \right)^{{1 \over m-1}}
  N\,.
 \end{equation} 
 Since any non-trivial pattern satisfies $\epsilon_{pat} > \epsilon_0$, 
 the absolute upper bound is $\delta Y_0 \sim N$. 
   In other words, 
 the memory burden stabilizes the system at the latest after its  {\it naive} 
 half-decay time, unless the memory pattern is offloaded beforehand. 
 
  As already discussed at length  in \cite{Gia1}, such offloading is possible if the system possesses another state of enhanced memory   
capacity for a different value of macro-parameter
 $Y_0 = N^{'}$ for which a different set of 
 modes $Y_{k^{'}}^{'}$ becomes gapless. 
 This can be modeled by the following Hamiltonian:   
    \begin{eqnarray}  \label{H444} 
 &&\hat{H}_Y  = \left ( 1 -  
 {\hat{Y}_0\over N}  \right )^m \sum_{k\neq 0} \epsilon_k \hat{Y}_k\, + \, \\ \nonumber 
 && +  \left ( 1 -  
 {\hat{Y}_0\over N^{'} }  \right )^m \sum_{k^{'}\neq 0} \epsilon_{k^{'}}^{'} \hat{Y}_{k^{'}}^{'}\, + \, ... \,,
  \end{eqnarray} 
  where $N^{'} <  N$ and ``$...$" includes mixing with the $b$-sector
  analogous to the terms in (\ref{H333}). 
  Correspondingly after changing the macro-parameter by  $\delta Y_0 = 
  N- N^{'}$,  a new set of memory modes $\hat{Y}_{k^{'}}^{'}$ becomes gapless while the old ones $\hat{Y}_k$ acquire non-zero gaps given by  
${\mathcal E}_k = \epsilon_k \left (  
 1-{ N^{'} \over N}  \right )^m$. \\

 In such a case the system could handle the memory 
 burden in two ways: 
 
 1) If $m$ is sufficiently large,  the memory 
 burden can be postponed until  $\delta Y_0$ becomes large. During 
 this time the pattern 
 can stay encoded in the $Y_k$-modes.
 
 2) Another option is that the memory burden gets eased by means of offloading the pattern 
 from $Y_k$-modes into $Y_{k^{'}}^{'}$-ones. 
 This makes  
 the decay of $Y_0$ possible.  Note that for the efficiency of 
 such a process, the  mixing between 
 $a_0$ and $b_0$ modes should be larger than $1/N$. 
Moreover, we remark that during this rewriting the pattern becomes  
scrambled meaning that in the new state the modes $Y_{k^{'}}^{'}$
become entangled with each other \cite{Gia1}.
   \\

 We do not know which of the above possibilities (or both) is realized in the de Sitter case. As we shall see, however, this does not change the 
 outcome over very long time scales.  Sooner or later the memory burden becomes unbearable.  \\

 It is important to understand that the memory burden effect does not 
 reduce to a statement that a system likes to be in a high entropy state,  
 although the two effects are related. 
 The entropy is a property of a macro-state
 whereas memory burden is a property of a particular micro-state 
 from a given macro-ensemble, {\it a priori} unrelated to the number 
 of fellow members in the ensemble.  However, 
 by simple combinatorics it is clear that for a system of micro-state 
 entropy $S$ the number of empty patterns is exponentially suppressed
 \cite{Gia1}. \\

\section{De Sitter and Inflation}

 We shall now apply the above knowledge to de Sitter.
  Let us first give a very general outline of  our reasoning. 
  
  We know that de Sitter has Gibbons-Hawking entropy $S$ and therefore 
  it must represent  a state of enhanced memory capacity achieved  
  for a critical value of a certain control parameter $Y_0$.   
  Whatever its precise origin is, we know that classically the value of this control parameter is set by
  $\Lambda$ which is a fixed parameter of the theory.   
  However,  the quantum evolution caused by Gibbons-Hawking evaporation 
  must lead to the change of $Y_0$ and thus to a subsequent  departure from the enhanced memory state. 
  This should result in a memory burden effect which must become 
 strong after a certain critical time. It is reasonable to assume 
 that this critical time must be bounded from above by the
 time during which the total energy radiated away via
 Gibbons-Hawking quanta becomes comparable to the 
 energy of the entire Hubble patch. The latter is given  by 
 $E_{dS} \sim SH$.  Thus, the effect must become strong at the latest after the total number of emitted quanta reaches $\sim S$. 
   We expect that the above  qualitative picture must be rather insensitive 
 to the details of the microscopic theory. 
 
Notice that at this point we do not make any extra assumption that ties the control parameter $Y_0$ to the energy of the system.  All we say is that when a memory-storing 
device loses half of its mass, it is reasonable to expect that it is 
pushed out of the original state of enhanced memory capacity. \\

   Next we shall try to be a bit more quantitative.   
  First of all, all the relevant quantities can be expressed in terms 
 of two parameters, the entropy $N=S$ and the Hubble scale  $H\sim\epsilon_0$. 
 Due to Gibbons-Hawking evaporation, after the time $t_Q = SH^{-1}$
a Hubble patch of size $H^{-1}$ would emit a number of 
quanta $N$ of order of its Gibbons-Hawking entropy $S$. 
The typical energy of individual quanta would be $\epsilon_0 \sim  H$. 
The integrated emitted energy would therefore be of the 
same order as the total vacuum energy contained within the Hubble patch 
which is equal to $E_{dS} \sim S\epsilon_0$.
This counting agrees with the physical meaning given to the time scale 
$t_Q = SH^{-1}$ in \cite{QB1,QB2,QB3} as the time after which 
the coherent state describing the de Sitter Hubble patch loses a fraction of order one of its constituents.  

These scalings  tell  us that in applying the general result of the previous section to de Sitter it should be treated as the special case $N=S$.  
 Next, since  de Sitter  has a micro-state entropy $S=N$, there must exist some (nearly-)gapless memory modes $Y_k$ of the number $S=N$ that support it. 
 Even without speculating where they come from, we can still gain some valuable information. For example, these modes can be labeled by quantum numbers 
that are symmetries of the de Sitter space in the classical limit. In order to have a level degeneracy $N$, these modes must  
belong to very high harmonics which gives an estimate 
$\epsilon_k \sim \sqrt{N} \epsilon_0$.  
Of course, this scaling also fully matches  the holographic 
counting \cite{Hol1} naively applied to de Sitter  which implies the
existence of $N$ Planck wavelength qubits.  

We can thus estimate 
that a typical {\it unactualized} energy of a memory pattern   
carried by a de Sitter patch is equal to
$\epsilon_{pat} \sim N^{3 \over 2} \epsilon_0 \sim E_{dS} \sqrt{N}$.  This scaling reveals how
incredibly efficient de Sitter's  memory storage 
is.  A pattern that with naive counting would exceed the 
energy of the entire de Sitter patch by a factor $\sqrt{N}$
is stored at the same cost as the empty pattern. Of course, this is nothing 
more than restating the fact of enormous micro-state degeneracy 
that underlies the Gibbons-Hawking entropy.   

So far, we have made no assumptions about 
the microscopic structure of de Sitter. We shall continue in the same spirit with the only obvious assumption that the loss of the quanta 
$\delta Y_0$ takes the de Sitter state away from the point of 
exact gaplessness of the $Y_k$-modes. In such a case a memory burden 
of the sort (\ref{mueff}) will be created. 

\begin{figure}
	\centering 
	\begin{subfigure}{0.3\textwidth}
		\includegraphics[width=\textwidth]{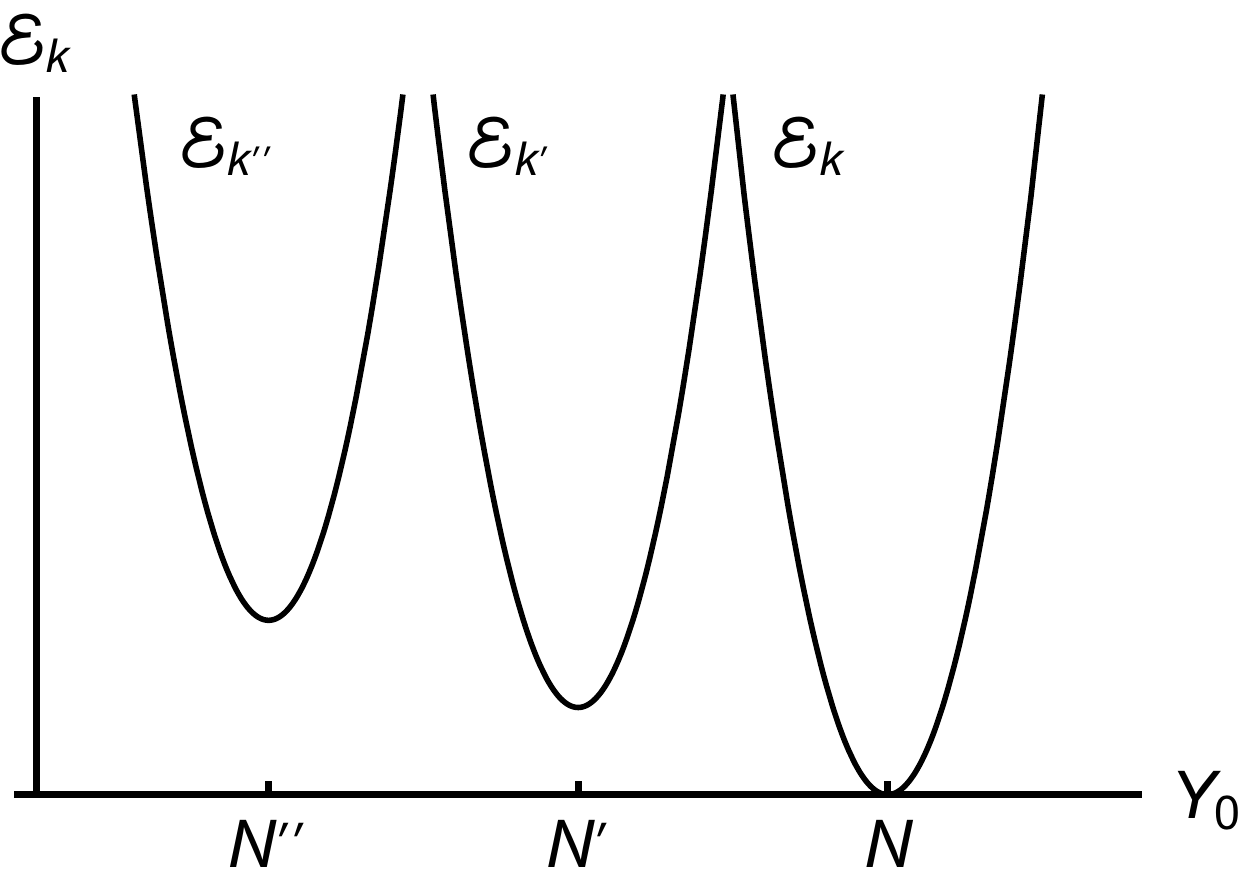}
		\caption{For small $m$, the minima are narrow.} 
		\label{sfig:biasedSmallM}
	\end{subfigure}
%	\hspace{0.1\textwidth}
	\begin{subfigure}{0.3\textwidth}
	\includegraphics[width=\textwidth]{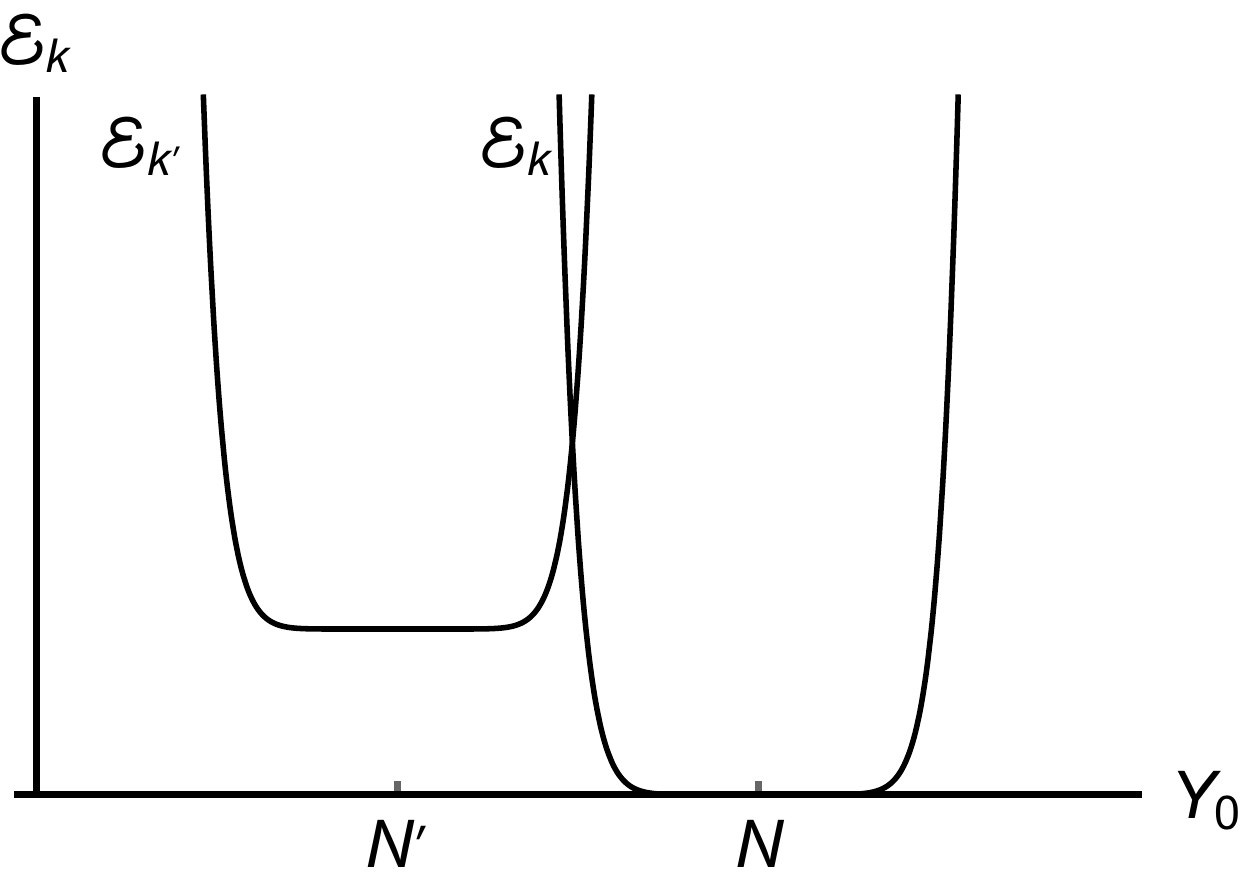}
		\caption{For large $m$, the minima are wide. }
		\label{sfig:biasedBigM}
	\end{subfigure}
	\caption{\raggedright Highly schematic plots (for even values of $m$) of the energy thresholds of the memory modes in a theory with cosmological constant. Only around a single value of $Y_0$, gapless modes emerge.}
	\label{fig:Biased}
\end{figure}

The following comment is very important. Since in de Sitter the critical number $N$ is set by the cosmological constant $\Lambda$, it 
represents a {\it fixed parameter} of the theory.  So even if de Sitter possesses 
information storing minima for other values of $Y_0=N'\neq N$, 
their energy must be an increasing function of  $|N-N'|$. 
For a better understanding, this can modeled by the following Hamiltonian: 
    \begin{eqnarray}  \label{H999} 
 &&\hat{H}_Y  = \left ( 1 -  
 {\hat{Y}_0\over N}  \right )^m \sum_{k\neq 0} \epsilon_k \hat{Y}_k\, + \, \\ \nonumber 
 && +  \left(\left ( 1 -  
 {\hat{Y}_0\over N^{'} }  \right )^m + \left(1 - {\hat{Y}_0\over N}
 \right)^l\right ) \sum_{k^{'}\neq 0} \epsilon_{k^{'}}^{'} 
  \hat{Y}_{k^{'}}^{'}\, + \, ...\,, 
  \end{eqnarray} 
where $l >0$. The resulting energy landscape is plotted in \fig \ref{fig:Biased}.

Now, assume that the two sets of modes 
$Y_{k^{'}}^{'},Y_k$ can carry an identical pattern, 
$Y_{k^{'}}^{'} = Y_k$.  
 Nevertheless, its energy costs in the state $Y_0=N'$ exceeds the 
 one in the state  $Y_0=N$ by  the amount $E_{pat} =\left(1 - {Y_0\over N}
  \right)^l \epsilon_{pat}$. So even if de Sitter will keep
  copying the pattern
  from one set of modes into another, the memory burden will
  increase steadily. 
 Since we do not fully specify the microscopic theory of de Sitter, 
 we cannot say what $m,l$ are. However, the universal constraint on the 
 parameters is that in the $N\rightarrow\infty$ limit only a single macro-state of enhanced 
 memory capacity must exist. This is necessary for matching the semi-classical description of de Sitter. Under these circumstances it is clear that the memory burden 
 should set in at the latest for $\delta Y_0 \sim N$. 
 After this point it strongly back reacts on the decay process. 
  Notice that this upper bound fully matches the quantum break-time 
 (\ref{Qt}). \\

The quantum breaking consequently is a result of two competing 
tendencies. On the one hand, the system depletes by losing the 
constituent  $Y_0$-modes into Gibbons-Hawking radiation
$X_0$.  This pushes  
the system away from criticality where the modes $Y_{k\neq 0}$ that carry the
{\it M-pattern}  are gapless. According to the energy gaps (\ref{eff}), the  {\it M-pattern} therefore  gradually becomes very costly in energy. 
On the other hand, it is impossible to offload this pattern into Gibbons-Hawking radiation due to the enormous level splitting 
between the critical modes $Y_k$  and their free counterparts
$X_k$.  This results into a back reaction 
 that makes Gibbons-Hawking emission unfavorable. 
 Thus, after a finite time the memory burden becomes unbearable and 
 the emission is tamed.  \\
  
  It is likely that quantum breaking of de Sitter is a signal of a fundamental quantum inconsistency of theories with positive {\it constant}  vacuum energy \cite{QB1,QB2,QB3} (for alternative views, see, e.g., \cite{Other, Polyakov}).   
 How does inflation change the story?  The slow-roll version of inflation \cite{slow} provides a new set of degrees of freedom coming from a scalar field, the inflaton 
$\hat{I}$,  which can coherently change the parameters and take 
the system out of the de Sitter phase before 
its quantum break-time. A microscopic description of this type of graceful exit was given in \cite{QB1}. 
 
   In this paper we shall not speculate on whether a sensible theory 
   can allow de Sitter cosmologies that extend  beyond the 
   quantum breaking  point.    
   Even without knowing this, we can still make some important physical conclusions.  
   The point is that, irrespectively of the consistency of the de Sitter state 
  beyond its quantum-break time, it is clear that within our Hubble patch the inflaton found a graceful exit beforehand.  
    Such a scenario has a very high likelihood because the semi-classical description of the late inflationary epoch shows no conflict with observations.  This gives us a chance to search for observable  imprints 
of the primordial memory burden. 
The read-out of this pattern becomes easier 
for a longer duration of inflation.  
 
 This may sound paradoxical but it is not.  This pattern represents a quantum hair stored in degrees of freedom that initially 
 were essentially gapless. It is not surprising that 
by quantum uncertainly a very long time is required for
decoding a pattern of such a narrow energy gap.   
      
   \section{De Sitter Versus Black Holes} 
\begin{figure}
 	\centering 
 	\begin{subfigure}{0.3\textwidth}
 		\includegraphics[width=\textwidth]{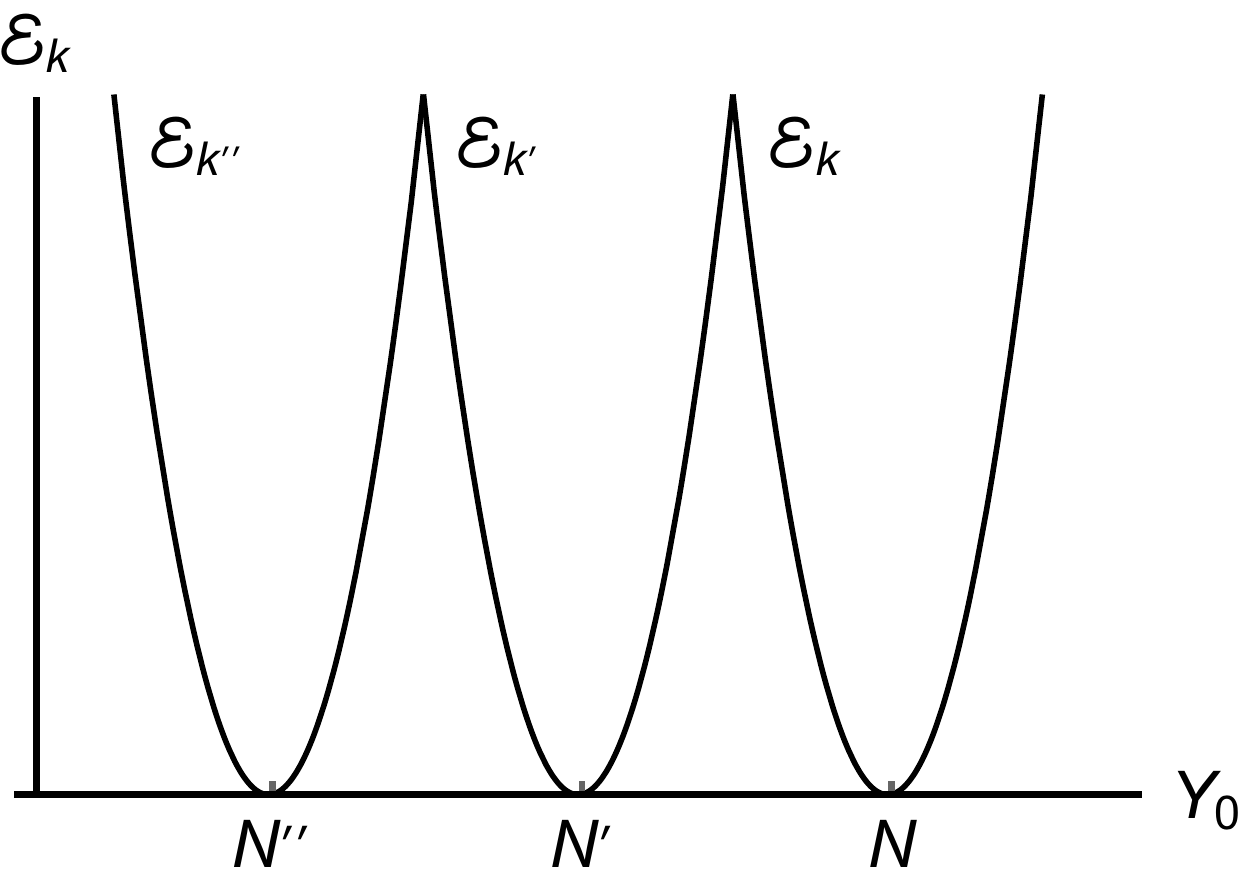}
 		\caption{Small $m$.} 
 		\label{sfig:smallM}
 	\end{subfigure}
 	%	\hspace{0.1\textwidth}
 	\begin{subfigure}{0.3\textwidth}
 		\includegraphics[width=\textwidth]{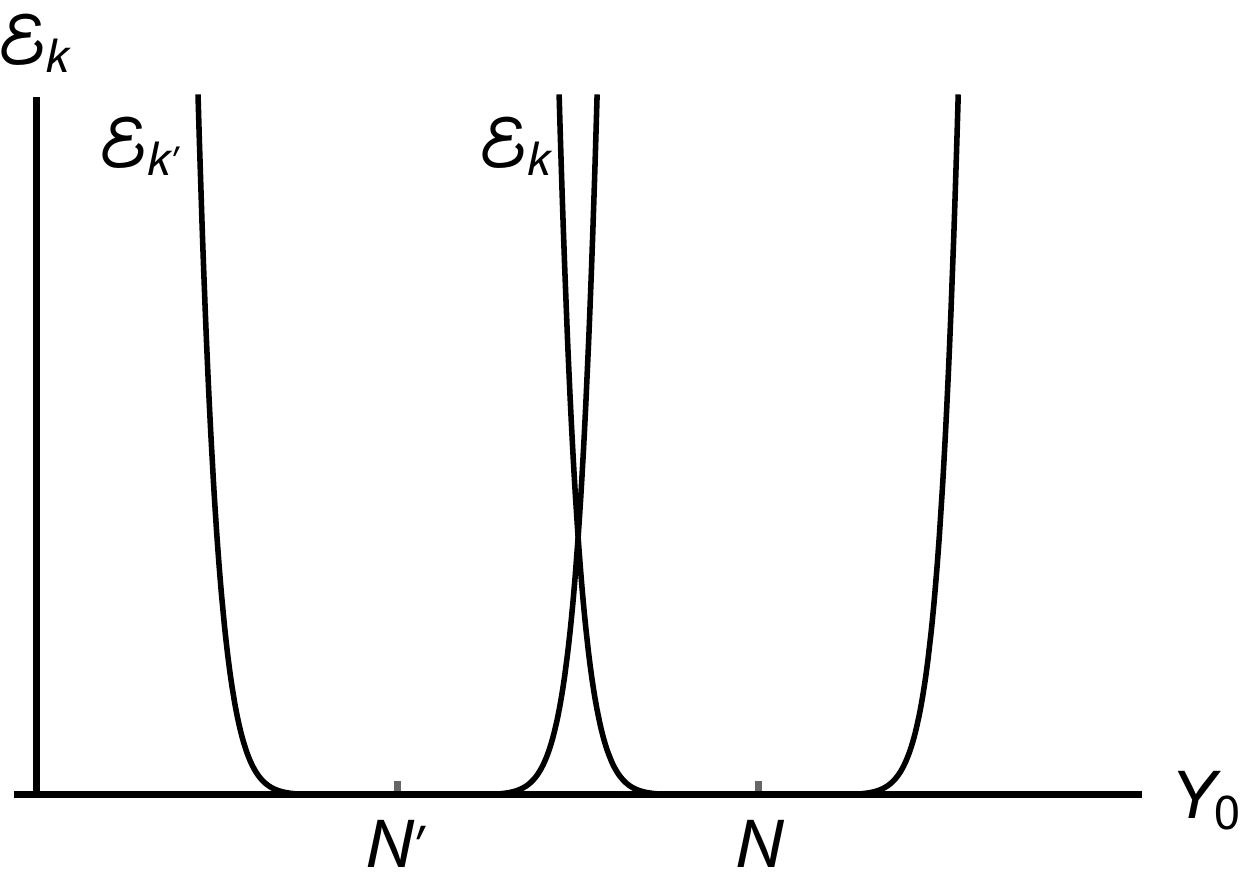}
 		\caption{Large $m$.}
 		\label{sfig:bBigM}
 	\end{subfigure}
 	\caption{\raggedright Highly schematic plots (for even values of $m$) of the energy thresholds of the memory modes for the case of black holes. Multiple minima exist, corresponding to different possible black hole masses.}
 	\label{fig:Unbiased}
 \end{figure}

 In this section we briefly ask whether by a similar analogy 
the memory overburden could 
lead to an inconsistency for black holes. 
The answer is no \cite{Gia1} and the reason is that unlike 
the cosmological constant, the black hole mass is not a parameter of the theory.  
 Correspondingly the same theory contains black holes  with different masses 
 and entropies.  Therefore, it possesses an entire family 
 of states of enhanced memory storage capacities with different 
 values of $N$. The corresponding energy landscape is plotted in \fig \ref{fig:Unbiased}.\\
 
The analysis of \cite{Gia1} indicates 
that the memory burden plays an important role 
in the Hawking evaporation process. However, this does not lead 
to any inconsistency since the system can resolve the conflict by 
constantly copying the stored pattern from one set of the gapless modes into another, during which the pattern gets scrambled.  
 In this sense there is 
a profound difference between the cases of black holes and de Sitter. 
The former can process the original memory pattern by rewriting 
and scrambling it, whereas de Sitter has much less flexibility 
since it is forever bound to a single enhanced memory state 
that is fixed by the cosmological constant $\Lambda$. 
 In this respect, we can say that 
a de Sitter Universe is a memory keeper, whereas black holes are both 
memory keepers and ``thinkers".

  \section{Outlook}
  In this paper we have pointed out a qualitatively new 
 mechanism by which primordial quantum information could have been 
 carried through the entire de Sitter epoch of the Universe's  
 history.

 First, we point out that de Sitter falls in a category 
 of states of maximal memory storage
 capacity due to its Gibbons-Hawking entropy. 
 Hence it must be subjected to the universal phenomenon of 
 {\it memory burden}  typical for such systems \cite{Gia1}.       
  Our point is that the primordial memory pattern ({\it M-pattern}),  which is carried 
  by the gapless modes responsible for the Gibbons-Hawking entropy, 
 cannot be erased by an inflationary time evolution. 
Instead, it creates a memory burden that grows with time and 
eventually clashes with the process of  Gibbons-Hawking evaporation after a finite time. This gives a new physical source of quantum breaking of de Sitter complementing the mechanism of \cite{QB1,QB2,QB3}. 

Obviously, inflation must have provided a graceful exit beforehand in our Hubble patch.  
  However, the longer inflation lasted, the higher the chances are to find an observational evidence of the primordial memory pattern. 
Answering the question how to detect it requires a separate investigation.   
 However, it is clear that the most interesting are the inflationary scenarios that end maximally close to the quantum break-time. In this case the memory burden due to the {\it M-pattern}  becomes close to unbearable 
and this will be imprinted in the primordial spectrum of density 
perturbations.  In this way, we could  -- at least in principle  -- read out the primordial quantum memories of our Universe.

\section*{Acknowledgements}
We thank  Cesar Gomez and Oleg Kaikov for valuable discussions and comments.  
This work was supported in part by the Humboldt Foundation under Humboldt Professorship Award and ERC Advanced Grant 339169 ``Selfcompletion".


\begin{thebibliography}{10}

\bibitem{Guth} A.~Guth, {\em The Inflationary Universe: A Possible Solution of Horizon and
	Flatness Problems}, 
\href{https://doi.org/10.1103/PhysRevD.23.347}{Phys. Rev. D23 (1981) 347}.

\bibitem{QB1}
G.~Dvali, C.~Gomez,
{\em Quantum Compositeness of Gravity: Black Holes, AdS and Inflation}, 
\href{https://doi.org/10.1088/1475-7516/2014/01/023}{JCAP 1401 (2014) 023}, 
\href{https://arxiv.org/abs/1312.4795}{arXiv:1312.4795 [hep-th]}.

\bibitem{QB2} G.~Dvali, C.~Gomez,
{\em Quantum Exclusion of Positive Cosmological Constant?}, 
\href{https://doi.org/10.1002/andp.201500216}{Annalen Phys. 528 (2016) 68}, 
\href{https://arxiv.org/abs/1412.8077}{arXiv:1412.8077 [hep-th]}.

\bibitem{QB3} G.~Dvali, C.~Gomez, S.~Zell,	
{\em Quantum Break-Time of de Sitter},
\href{https://doi.org/10.1088/1475-7516/2017/06/028}{JCAP 1706 (2017) 028,}
\href{http://arxiv.org/abs/1701.08776}{arXiv:1701.08776 [hep-th]}. 

\bibitem{GH} 
G.W.~Gibbons, S.W.~Hawking, 
{\em Cosmological Event Horizons, Thermodynamics, and Particle Creation},  
\href{https://doi.org/10.1103/PhysRevD.15.2738}{Phys. Rev. D15 (1977) 2738}.   

\bibitem{qbreaking} 	
G.~Dvali, D.~ Flassig, C.~ Gomez, A.~ Pritzel, N.~ Wintergerst, 
{\em Scrambling in the Black Hole Portrait},
\href{https://doi.org/10.1103/PhysRevD.88.124041}{Phys. Rev. D88 (2013)  124041}, 
\href{https://arxiv.org/abs/1307.3458}{arXiv:1307.3458 [hep-th]}. 

\bibitem{IR}
U.~H.~Danielsson, D.~Domert, M.~E.~Olsson,
{\em Miracles and complementarity in de Sitter space},
\href{https://doi.org/10.1103/PhysRevD.68.083508}{Phys.\ Rev.\ D68 (2003) 083508},
\href{https://arxiv.org/abs/hep-th/0210198}{arXiv:hep-th/0210198};

S.~B.~Giddings and M.~S.~Sloth,
{\em Semiclassical relations and IR effects in de Sitter and slow-roll space-times},
\href{https://doi.org/10.1088/1475-7516/2011/01/023}{JCAP 1101 (2011) 023}, 
\href{https://arxiv.org/abs/1005.1056}{arXiv:1005.1056 [hep-th]};

R.~Z.~Ferreira, M.~Sandora, and M.~S.~Sloth, 
{\em Asymptotic Symmetries in de
	Sitter and Inflationary Spacetimes}, 
\href{https://doi.org/10.1088/1475-7516/2017/04/033}{JCAP 1704 (2017) 033},
\href{https://arxiv.org/abs/1609.06318}{arXiv:1609.06318 [hep-th]};

T.~Markkanen,  
{\em De Sitter Stability and Coarse Graining}, 
\href{https://doi.org/10.1140/epjc/s10052-018-5575-9}{Eur. Phys. J. C78 (2018) 97}, 
\href{https://arxiv.org/abs/1703.06898}{arXiv:1703.06898 [gr-qc]}.

R.~Brustein,  Private communications, unpublished.  


\bibitem{measure} 
N.~Arkani-Hamed, S.~Dubovsky, A.~Nicolis, E.~Trincherini, and
G.~Villadoro, {\em A Measure of de Sitter Entropy and Eternal Inflation},
\href{https://doi.org/10.1088/1126-6708/2007/05/055}{JHEP 05 (2007) 055},
\href{https://arxiv.org/abs/0704.1814}{arXiv:hep-th/07041814}.

\bibitem{DGZ} 
G.~Dvali, C.~Gomez,  
{\em On Exclusion of Positive Cosmological Constant},  
\href{https://arxiv.org/abs/1806.10877}{arXiv:1806.10877 [hep-th]};

G.~Dvali, C.~Gomez, S.~Zell, 
{\em Quantum Breaking Bound on de Sitter and Swampland},  
\href{https://arxiv.org/abs/1810.11002}{arXiv:1810.11002 [hep-th]}.

\bibitem{swamp} 
G.~Obied, H.~Ooguri, L.~Spodyneiko, C.~Vafa,	
{\em De Sitter Space and the Swampland}, 
\href{https://arxiv.org/abs/1806.08362}{arXiv:1806.08362 [hep-th]}.

For a later refined version, see, 

H.~Ooguri, E.~Palti, G.~Shiu, C.~Vafa,
{\em   Distance and de Sitter Conjectures on the Swampland},
\href{https://doi.org/10.1016/j.physletb.2018.11.018}{Phys. Lett. B788 (2019) 180},
\href{https://arxiv.org/abs/1810.05506}{arXiv:1810.05506 [hep-th]}.

\bibitem{Gia1}
G.~Dvali, {\em A Microscopic Model of Holography: Survival by the Burden of Memory}, 
\href{https://arxiv.org/abs/1810.02336}{arXiv:1810.02336 [hep-th]}. 

\bibitem{Bek}
J.~D. Bekenstein, {\em Black holes and entropy\/},
\href{http://dx.doi.org/10.1103/PhysRevD.7.2333}{Phys. Rev. D7 (1973) 2333}.

\bibitem{BBound}
J.~D. Bekenstein, {\em Universal upper bound on the entropy-to-energy ratio for bounded systems\/},
\href{http://dx.doi.org/10.1103/PhysRevD.23.287}{Phys.
	Rev. D23 (1981) 287}. 

\bibitem{Gia2} G.~Dvali, 
{\em Area Law Micro-State Entropy from Criticality and Spherical Symmetry}, 
\href{https://doi.org/10.1103/PhysRevD.97.105005}{Phys.Rev. D97 (2018), 105005},  
\href{https://arxiv.org/abs/1712.02233}{arXiv:1712.02233 [hep-th]}; 

{\em Black Holes as Brains: Neural Networks with Area Law Entropy}, 
\href{https://doi.org/10.1002/prop.201800007}{Fortsch. Phys. 66 (2018) 040007}, 
\href{https://arxiv.org/abs/1801.03918}{arXiv:1801.03918 [hep-th]};

{\em  Critically excited states with enhanced memory and pattern recognition capacities in quantum brain networks: Lesson from black holes},
\href{https://arxiv.org/abs/1711.09079}{arXiv:1711.09079 [quant-ph]}.

\bibitem{Atoms}	
G.~Dvali,  M.~ Michel, S.~ Zell,  
{\em Finding Critical States of Enhanced Memory Capacity in Attractive Cold Bosons},  
\href{https://arxiv.org/abs/1805.10292}{arXiv:1805.10292 [quant-ph]}. 

\bibitem{Hol1} 
G.'t Hooft, {\em Dimensional reduction in quantum gravity},
\href{https://arxiv.org/abs/gr-qc/9310026}{arXiv:gr-qc/9310026};

L. Susskind,
{\em The World as a Hologram}, 
\href{https://doi.org/10.1063/1.531249}{J. Math. Phys. 36 (1995) 6377},
\href{https://arxiv.org/abs/hep-th/9409089}{arXiv:hep-th/9409089}.

\bibitem{Other} 
T.~Banks and W.~Fischler, {\em An upper bound on the number of e-foldings},
\href{https://arxiv.org/abs/astro-ph/0307459}{arXiv:astro-ph/0307459};

R.~Bousso, B.~Freivogel and I.S.~Yang,
{\em Eternal inflation: The inside story},
\href{https://doi.org/10.1103/PhysRevD.74.103516}{Phys. Rev. D 74 (2006) 103516},
\href{https://arxiv.org/abs/hep-th/0606114}{arXiv:hep-th/0606114};

\mbox{T.~Banks}, {\em Holographic Inflation Revised}, 
\href{https://arxiv.org/abs/1501.01686}{arXiv:1501.01686 [hep-th]}.

\bibitem{Polyakov} 
It is of fundamental importance not to confuse  the quantum breaking phenomenon with a possible instability of the type suggested in   

A.M.~Polyakov,
{\em De Sitter Space and Eternity},
\href{https://doi.org/10.1016/j.nuclphysb.2008.01.002}{Nucl. Phys. B797 (2008) 199},
\href{https://arxiv.org/abs/0709.2899}{arXiv: 0709.2899 [hep-th]};

{\em Infrared instability of the de Sitter space},
\href{https://arxiv.org/abs/1209.4135}{arXiv:1209.4135 [hep-th]}.
Quantum breaking \cite{QB1,QB2,QB3}  is {\it not } accompanied by any Lyapunov exponent  that could potentially provide a graceful exit from the problem and this is the source of inconsistency. 

\bibitem{slow} See, e.g., A.D.~Linde, {\em Particle Physics and Inflationary Cosmology}, Harwood
Academic, Switzerland (1990); 
E.W.~Kolb and M.S.~Turner, {\em The Early
Universe}, Addison Wesley (1990).
























%\bibitem{class} G.~Dvali,  Classicalization Clearly: Quantum Transition into States of Maximal Memory Storage Capacity,  arXiv:1804.06154 [hep-th] 

%\bibitem{bogoliubov}
%N.~Bogoliubov,
%{\em On the theory of superfluidity},  J. Phys.  11 no.~1, 23.


%\bibitem{NP}
%G.~Dvali and C.~Gomez,
%{\em Black Hole's Quantum N-Portrait},
%\href{https://doi.org/10.1002/prop.201300001}{Fortsch. Phys.  61 (2013) 742}
%\href{https://arxiv.org/abs/1112.3359}{arXiv:1112.3359 [hep-th]}.  

%\bibitem{DG}
%G.~Dvali and C.~Gomez,
%{\em Black Holes as Critical Point of Quantum Phase Transition},
%\href{https://doi.org/10.1140/epjc/s10052-014-2752-3}{Eur. Phys. J. C  74 (2014)  2752},
%\href{https://arxiv.org/abs/1207.4059}{arXiv:1207.4059 [hep-th]}.	






%\bibitem{3Dnonperiod}
% 	G.~Dvali and L.Espinoza, in preparation.

%\bibitem{fast}P.~Hayden and J.~Preskill, Black holes as mirrors: Quantum
%information in random subsystems, JHEP 0709 (2007) 120,
%arXiv:0708.4025 [hep-th].
%
%Y.~Sekino and L.~Susskind, Fast Scramblers, JHEP 0810 (2008) 065,
%arXiv:0808.2096 [hep-th].



% For the model with more gapless modes, see: 
% G.~Dvali, A.~Franca, C.~Gomez, and N.~Wintergerst, Nambu-Goldstone
%		Effective Theory of Information at Quantum Criticality,
%Phys. Rev. D92 (2015) 125002, arXiv:1507.02948 [hep-th].

%\bibitem{numbers} 
%G.~Dvali, L.~Eisemann, M.~Michel, S.~Zell,  work in progress. 

%\bibitem{atoms} I.~Bloch, J.~Dalibard, and S.~Nascimbène, Quantum simulations with ultracold
%quantum gases, Nature Phys. 8 (2012), 267;
%I.~Bloch, J.~Dalibard, W.~Zwerger, Many-Body Physics with Ultracold Gases,
%Rev. Mod. Phys. 80 (2008), 885, arXiv:0704.3011 [cond-mat.other].















	
\end{thebibliography}
\end{document}